\documentclass[conference]{IEEEtran}
\IEEEoverridecommandlockouts
\usepackage{cite}
\usepackage{amsmath,amssymb,amsfonts}
\usepackage{algorithmic}
\usepackage{graphicx}
\usepackage{textcomp}
\usepackage{xcolor}
\usepackage{setspace}
\usepackage{booktabs}

\def\BibTeX{{\rm B\kern-.05em{\sc i\kern-.025em b}\kern-.08em
    T\kern-.1667em\lower.7ex\hbox{E}\kern-.125emX}}
\begin{document}
\fontsize{9pt}{11.4}\selectfont
\title{\textbf{Robust Target Speaker Direction of Arrival Estimation }\\
}

\author{\IEEEauthorblockN{ Zixuan Li, Shulin He, Xueliang Zhang}
\IEEEauthorblockA{\textit{College of Computer Science} \\
\textit{Inner Mongolia University}\\
Hohhot, China \\
   cslzx@mail.imu.edu.cn, heshulin@mail.imu.edu.cn, cszxl@imu.edu.cn
}

}

\maketitle

\begin{abstract}
In multi-speaker environments the direction of arrival (DOA) of a target speaker is key for improving speech clarity and extracting target speaker's voice. However, traditional DOA estimation methods often struggle in the presence of noise, reverberation, and particularly when competing speakers are present. To address these challenges, we propose RTS-DOA, a robust real-time DOA estimation system. This system innovatively uses the registered speech of the target speaker as a reference and leverages full-band and sub-band spectral information from a microphone array to estimate the DOA of the target speaker's voice. Specifically, the system comprises a speech enhancement module for initially improving speech quality, a spatial module for learning spatial information, and a speaker module for extracting voiceprint features. Experimental results on the LibriSpeech dataset demonstrate that our RTS-DOA system effectively tackles multi-speaker scenarios and established new optimal benchmarks.

% our research presents a robust, real-time DOA estimation framework that utilizes both full-band and sub-band spectral information from microphone arrays, along with registered speech from the target speaker. This innovative approach significantly improves the accuracy of locating the target speaker's position. Our model, rigorously tested, not only demonstrates a streamlined design conducive to real-time processing but also significantly outperforms existing studies. Notably, it maintains exceptional performance in conditions of high reverberation and low signal-to-noise ratio, representing a notable leap forward in DOA estimation technology.

\end{abstract}


\section{Introduction}
Target speaker direction of arrival (DOA) estimation aims to accurately locate the sound source of a target speaker in multi-speaker environments. In complex environments such as conference rooms and public spaces, accurate DOA estimation plays a critical role in enabling sound source separation, thereby enhancing the accuracy and efficiency of automatic speech recognition (ASR) systems. Accurate DOA estimation for the target speaker is key in noisy multi-speaker environments for identifying the main sound source and suppressing background noise. Our previous work \cite{he20243sicassp} has proven its importance.
% Previous work \footnote{S. He, H. Li, Y. Yang, F. Chen, X. Zhang et al., “3s-tse: Effi-cient three-stage target speaker extraction for real-time and low-resource applications,” arXiv preprint arXiv:2312.10979, 2023.\label{3s}} has proven its effectiveness.
% , where we initially estimated the target speaker's DOA using a multi-microphone array. We then used the Generalized Sidelobe Canceller (GSC) to separate the target speaker's voice from the mixed signal, followed by post-processing to refine and output the final separated audio. This approach significantly improved the short-time objective intelligibility (STOI) \cite{taal2010short} by $ 17.3 \% $ on the Librispeech dataset while maintaining a compact model with only 0.19M parameters.

In the field of DOA estimation, many classical signal processing techniques have been extensively studied and applied. Broadly, these techniques are categorized into: 1) Subspace methods like Multiple Signal Classification (MUSIC) \cite{schmidt1986multiple} and the estimation of signal parameters via rotational invariance technique (ESPRIT) \cite{roy1989esprit}, 2) Approaches based on Time Difference of Arrival (TDOA), which utilize generalized cross-correlation (GCC) methods \cite{huang2001real,knapp1976generalized}, 3) Methods focusing on signal synchronization, for example, Steered Response Power with Phase Transform (SRP-PHAT) \cite{brandstein1997robust} and Multichannel Cross Correlation Coefficient (MCCC) \cite{benesty2004time}, and 4) Techniques that are model-based, such as the maximum likelihood approach \cite{DBLP:journals/tsp/StoicaS90}. However, these conventional methods face challenges such as unrealistic assumptions about signal/noise models\cite{tho2014robust}, high computational load and limited effectiveness in low signal to noise ratios (SNR) or highly reverberant conditions \cite{benesty2008microphone}.

Deep neural networks (DNNs) have recently significantly enhanced speech processing and Direction of Arrival (DOA) estimation, particularly in the context of multi-microphone speech signal processing. \cite{luo2019conv,wang2021multi,luo2020end}, highlighting the significant potential of multi-channel speech in DNN-based speech signal processing. \cite{he20243sicassp} first integrated multi-channel speech and voiceprint features to estimate the target speaker DOA, subsequently using the Generalized Sidelobe Canceller (GSC) \cite{griffiths1982alternative}  to extract the target speaker's voice from mixed signals, significantly improving the Short-Time Objective Intelligibility (STOI) \cite{taal2010short}. However, the accuracy of DOA estimation was moderate, with performance drastically declining in noisy environments.
\cite{takeda2016sound} utilized a DNN with seven hidden layers to predict DOA using eigenvectors of correlation matrices as input, though their method is sensitive to reverberation. \cite{xiao2015learning} achieved noise robustness by employing GCC features in a perceptron network. \cite{chakrabarty2017broadband}'s CNN approach, which uses STFT phase components, demonstrated resilience to noise and microphone variations. \cite{li2018online} combined CNN and LSTM for online DOA estimation in challenging conditions, while \cite{mack2020signal,mack2022signal} incorporated attention mechanisms into DOA estimation. Although these methods partially address noise and reverberation challenges, they typically assume a stationary target speaker and struggle with scenarios involving multiple interfering speakers, as they cannot effectively distinguish the primary sound source within mixed speech.

% In recent years, many studies have employed microphone arrays for multi-channel speech separation and achieved impressive results \cite{luo2019conv,wang2021multi,luo2020end}, revealing the powerful potential of microphone arrays in speech signal processing. Furtherfore, a number of studies have suggested the use of both full-band/cross-band and sub-band/narrow-band data in a segregated manner, followed by their integration. This approach has demonstrated notable success in tasks such as voice separation and enhancement, exemplified by developments like TFGridNet\cite{wang2023tf} and SpatialNet\cite{quan2024spatialnet}.

To address the challenges of accurately estimating the target speaker DOA in complex scenarios with noise, reverberation, interfering speakers, and target speaker movement, thereby improving outcomes for downstream tasks, this paper introduces RTS-DOA, a robust, deep neural network-based model for target speaker DOA estimation using multi-channel speech. Specifically, the speech enhancement module in the model preliminarily improves the signal quality, while the speaker feature module extracts local features. The Spatial module models the spatial information using full-band sub-band information, processed speech, and local features, and finally outputs the estimated DOA. Compared to the baseline, the accuracy of DOA prediction improved by about $30\%$, while maintaining a compact model with only 0.12M parameters. 
% Furthermore, we applied the full-band sub-band network to the DOA estimation task, significantly improving the precision without increasing computational resource consumption.

\section{System overview}
\label{System Overview}

\begin{figure}[t]
  \centering
  \includegraphics[width=\linewidth]{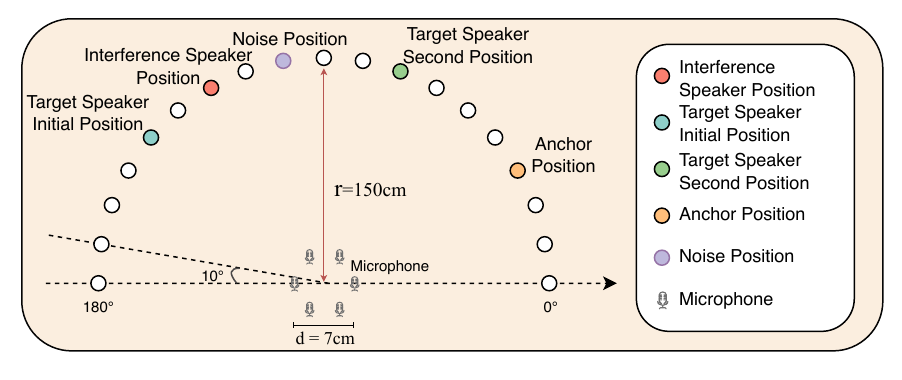}
  \caption{Structure of the six-channel microphone array. For brevity, the figure presents the source signal distribution from 0° to 180°. The unseen distribution from 180° to 360° is a mirror reflection of the shown pattern, ensuring symmetry.}
  \label{fig:SystemOverview}
\end{figure}

As shown in Figure \ref{fig:SystemOverview}, RTS-DOA system comprises a circular array of six omnidirectional microphones, denoted as $M_{1}, \dots, M_{6}$. We train the system using 36 source locations, spaced evenly from 0° to 360° at 10° intervals, with each location 1.5 m from the center of the microphone array. For each training example, we randomly select five distinct locations from the 36 source locations. These five locations represent the initial position, second position, interfering position, anchor position for the target speaker and the location of the point source noise. To simulate the movement of the target speaker, we randomly shift the speaker from the initial position to the second position at a randomly chosen time point during each utterance simulation.

\section{Method}

\begin{figure*}[t]
  \centering
  \includegraphics[width=0.95\textwidth,keepaspectratio]{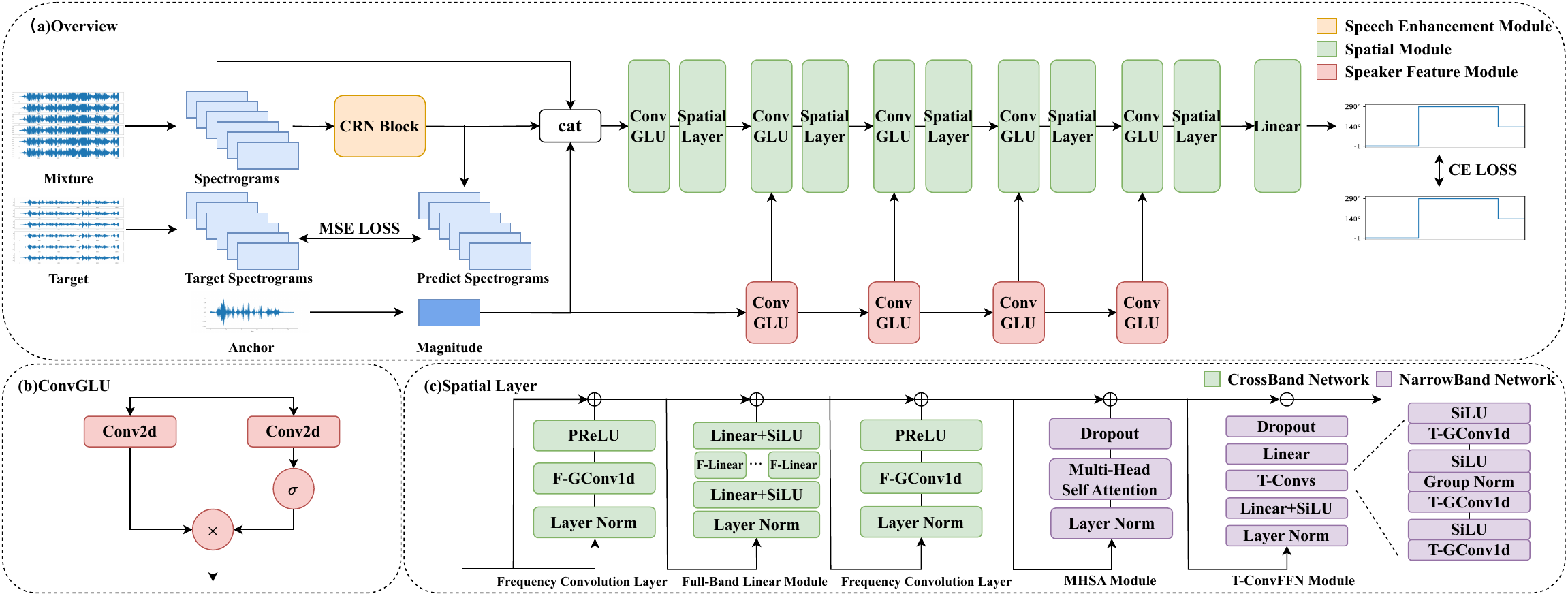}
  \caption{(a) The overall structure of the RTS-DOA system. (b) The structure of ConvGLU, where $\sigma$ represents the Sigmoid activation function and $\odot$ denotes element-wise multiplication. (c) The detailed structure of the Spatial Layer.
  \vspace{-0.4cm}
  }
  \label{fig:ModelFramework}
\end{figure*}

The overall structure of RTS-DOA system is shown in Figure \ref{fig:ModelFramework}(a), which mainly consists of three parts: (1) Speech Enhancement Module, (2) Spatial Module, (3) Speaker Feature Module.

First, the complex spectral of the multi-channel microphones is taken as the input for the Speech Enhancement Module, where the complex spectral is obtained by concatenating the real and imaginary parts of each Time-Frequency (T-F) bin. Subsequently, the output of the Speech Enhancement Module is concatenated with the magnitude spectrum of the anchor and then used as the input for the Spatial Module. At the same time, the magnitude spectrum of the anchor serves as the input for the Speaker Feature Module. Local features are obtained through the Speaker Feature Module and are concatenated at the corresponding layer in the Spatial Module. Afterwards, the DOA Estimation Module outputs the estimated DOA.

Next, we will provide a detailed introduction to the three main components of the proposed system.

\subsection{Speech Enhancement Module}

The Speech Enhancement Module uses the Convolutional Recurrent Neural Network (CRN) \cite{tan2018convolutional}, featuring an encoder-decoder structure with five convolutional and deconvolutional layers, respectively, and two LSTM layers between them for temporal modeling. The CRN processes the mixed speech's complex spectrum, producing noise-reduced multi-channel speech for DOA estimation. Its output is refined through Mean Squared Error (MSE) loss comparison with the clean speech's complex spectrum, optimizing the module's learning.

\subsection{Spatial Module}

\label{doa}

The Spatial Module consists of five interwoven Convolutional Gated Linear Units (ConvGLU) \cite{tan2019learning} blocks and Spatial Layers \cite{quan2024spatialnet}, along with a linear layer for predicting the classification results. The complex spectral of the multi-channel microphones and the output of the Speech Enhancement Module are concatenated with the magnitude spectrum of the anchor to serve as the input. The local features obtained from the Speaker Feature Module(This will be introduced in \ref{subsection:Speaker Feature Network}) are concatenated with the corresponding hidden states and then processed through the respective ConvGLU and Spatial Layers. Finally, the predicted DOA is obtained through the linear layer.
\subsubsection{ConvGLU}
The structure of the ConvGLU is shown in Figure \ref{fig:ModelFramework}(b). It extended the Convolutional Recurrent Neural Network by introducing a gating mechanism to control the information flow within the network, achieving good results in modeling complex spectral\cite{tan2019learning}. In this system, the same ConvGLU structure is used to extract local patterns from the complex spectral and reduce the feature resolution. The computational process can be represented as follows:
\begin{equation}
y = tanh( x * W_{1} + b_{1}) \odot \sigma ( x * W_{2} + b_{2})\label{con:ConvGLU}
\end{equation}
\(W\) and \(b\) represent the kernel and bias, respectively, \(\sigma\) represents the sigmoid function. The symbols \(*\) and \(\odot\) denote convolution operation and element-wise multiplication, respectively.

\subsubsection{Spatial Layer}

The Spatial Layer has the same structure to that described in \cite{quan2024spatialnet}, designed to learn complex spatial information. It consists of a CrossBand Network and a NarrowBand Network, as shown in Figure \ref{fig:ModelFramework}(c).

The CrossBand Network is designed to learn cross-band information, consisting of two frequency convolution layers and a full-band linear module. It processes each time frame independently, and all time frames use the same weights. The frequency convolution layer is used to model the correlation between adjacent frequencies, where F-GConv1d represents one group convolution along the frequency axis. The full-band linear module models the entire frequency spectrum to mitigate the impact of interfering speakers, noise, and reverberation, which narrowband signals cannot accurately capture in terms of spatial information. The first Linear layer is employed to change the dimensionality of channel to $H$, and the second Linear layer is used to restore the dimensions. Each F-Linear operates independently on the channel dimension.

The NarrowBand Network is designed to capitalize on the rich temporal information present in narrow-band signals. The NarrowBand Network processes each frequency independently, and all frequencies share weights. The NarrowBand Network is composed of one multi-head self-attention (MHSA) \cite{vaswani2017attention}  module and one time-convolutional feed forward network (T-ConvFFN). The MHSA is designed to utilize the self-attention mechanism to identify and segregate components belonging to the same or different speakers within an audio signal. T-ConvFFN is used to replace the feedforward network in the Transformer \cite{vaswani2017attention} architecture. The first Linear layer is employed to change the dimensionality of channel to $H^{\prime}$, and the second Linear layer is used to restore the dimensions.

\subsection{Speaker Feature Module}\label{subsection:Speaker Feature Network}

The Speaker Feature Module has the exact same structure as the ConvGLU in the Speech Enhancement Module. ConvGLU encodes the magnitude spectrum of the anchor, and the average value of the encoded magnitude spectrum for each frame is computed as the speaker feature. Speaker feature are concatenated, layer-by-layer, with the matching ConvGLU layers in the DOA module for each time frame \cite{he2020speakerfilter}, as depicted in Figure \ref{fig:ModelFramework}(a).

\subsection{Loss function}

The three main components of the proposed system, namely the Speech Enhancement Module, Spatial Module, and Speaker Feature Module, are trained jointly. The loss function consists of two parts:

\begin{equation}
\mathcal{L} = \mathcal{L}_{MSE} + \mathcal{L}_{CE} \label{con:loss}
\end{equation}

$ \mathcal{L}_{MSE} $,which represents the Mean Squared Error (MSE) loss between the output of the Speech Enhancement Module and the complex spectral of the clean speech, can be expressed as:

\begin{equation}
\mathcal{L}_{MSE} = \frac{1}{N} \sum_{n=1}^{N} \left( (\text{R}(Y_n) - \text{R}(\hat{Y}_n))^2 + (\text{I}(Y_n) - \text{I}(\hat{Y}_n))^2 \right) \label{con:mse_loss}
\end{equation}
Where \(\text{R}(Y_n)\) and \(\text{I}(Y_n)\) respectively represent the real and imaginary parts of the target complex spectrum at the \(n\)th frequency point, and \(\text{R}(\hat{Y}_n)\) and \(\text{I}(\hat{Y}_n)\) respectively represent the real and imaginary parts of the complex spectrum output by the Speech Enhancement Module at the \(n\)th frequency point.

\(\mathcal{L}_{CE}\) is the cross-entropy loss, which compares the true class labels with the predicted probabilities derived from the softmax function.

\section{Experience}

\begin{table*}[!htbp]
    \caption{Comparative Analysis of Voice Disambiguation Efficiency (VDE) and Accuracy Rate (AR) Across Various Signal-to-Noise Ratios (SIR) for Different Models}
    \label{tab:noisy}
    \scalebox{0.89}{
    \begin{tabular}{cccccccccccccccc}
    \toprule
    ID & SIR             & \multicolumn{2}{c}{-5} & \multicolumn{2}{c}{-2} & \multicolumn{2}{c}{0} & \multicolumn{2}{c}{2} & \multicolumn{2}{c}{5} & \multicolumn{2}{c}{AVG} \\
    \cmidrule(lr){3-4}\cmidrule(lr){5-6}\cmidrule(lr){7-8}\cmidrule(lr){9-10}\cmidrule(lr){11-12}\cmidrule(lr){13-14}
     & Metrics         & VDE $\downarrow$ & AR $\uparrow$ & VDE $\downarrow$ & AR $\uparrow$ & VDE $\downarrow$ & AR $\uparrow$ & VDE $\downarrow$ & AR $\uparrow$ & VDE $\downarrow$ & AR $\uparrow$ & VDE $\downarrow$ & AR $\uparrow$ \\ 
    \hline
    0 & TS-DOA        & 0.1657 & 0.5015 & 0.1369 & 0.5497 & 0.1464 & 0.5923 & 0.0981 & 0.6481 & 0.1007 & 0.6624 & 0.1285 & 0.6030 \\
    \hline
    1 & w/o Enhancement & 0.1643 & 0.6867 & 0.1503 & 0.7015 & 0.1457 & 0.7582 & 0.1364 & 0.7210 & 0.1285 & 0.7881 & 0.1497 & 0.7117 \\
    2 & w/o Local   & 0.1782 & 0.6384 & 0.1717 & 0.7225 & 0.1487 & 0.7102 & 0.1430 & 0.7443 & 0.1230 & 0.7907 & 0.1487 & 0.7093 \\
    3 & w global  & 0.1558 & 0.7723 & 0.1232 & 0.7838 & 0.1174 & 0.8111 & 0.1113 & 0.8566 & 0.0999 & 0.8730 & 0.1131 & 0.8220 \\
    4 & RTS-DOA(Mag)   & 0.1899 & 0.5328 & 0.1689 & 0.0273 & 0.1723 & 0.6790 & 0.1422 & 0.6926 & 0.1321 & 0.7476 & 0.1559 & 0.6642 \\
    \hline
    5 & \textbf{RTS-DOA}        & \textbf{0.1236} & \textbf{0.8940} & \textbf{0.1019} & \textbf{0.8837} & \textbf{0.0843} & \textbf{0.9166} & \textbf{0.0830} & \textbf{0.9136} & \textbf{0.0650} & \textbf{0.9364} & \textbf{0.0896} & \textbf{0.8940} \\
    6 & \textbf{RTS-DOA(Large)} & \textbf{0.1089} & \textbf{0.8632} & \textbf{0.0830} & \textbf{0.9061} & \textbf{0.0758} & \textbf{0.9394} & \textbf{0.0698} & \textbf{0.9355} & \textbf{0.0632} & \textbf{0.9486} & \textbf{0.0769} & \textbf{0.9167} \\
    \bottomrule
    \end{tabular}
    }
    \vspace{-0.4cm}
\end{table*}

\begin{table}
  \caption{Performance Comparison of Different Models in Terms of Parameters and Computational Cost}
  \label{tab:size}
  \centering
  \begin{tabular}{ c c c }
    \toprule
    \multicolumn{1}{c}{\textbf{Metrics}} & 
    \multicolumn{1}{c}{\textbf{Params(M)}} &
     \multicolumn{1}{c}{\textbf{MACs(G/s)}} \\
    \midrule
        TS-DOA & 0.164 & 0.147 \\
        RTS-DOA & 0.121 & 0.187 \\
        RTS-DOA(Large) & 1.334 & 0.738 \\
    \bottomrule
    \vspace{-0.6cm}
  \end{tabular}
\end{table}

\subsection{Dataset and setup}
For the training and testing data, we generated six-channel mixed speech by simulating a reverberant room environment, with target and interfering speakers randomly placed within the room. The speech data is sourced from the LibriSpeech corpus \cite{panayotov2015librispeech}, and the noise data is from the DNS Challenge \cite{dubey2022icassp}. Both the training and test sets consist of mixtures of target speaker speech, interfering speaker speech, and background noise. The room dimensions are 
5m$\times$6m$\times$3m, and the Room Impulse Responses (RIR) were generated using the Image method \cite{allen1979image}. The reverberation time (T60) was randomly chosen from {0.2, 0.3, 0.4, 0.5, 0.6, 0.7} seconds. The Signal-to-Interference Ratio (SIR) between target and interfering speech was randomly sampled from -5 dB to 5 dB in 1 dB steps, while the Signal-to-Noise Ratio (SNR) between target speech and noise was sampled within the same range and step size. Each sample is paired with a reference speech from LibriSpeech randomly sampled from the target speaker. The training set contains 20,000 samples, the evaluation set 600 samples, and the test set 1,000 samples, ensuring that speakers in the test set are not present in the training data. The shortest speech duration is 3 seconds, and the longest is 16 seconds, with a sampling rate of 16 kHz.

In the speech enhancement module, we adjusted the number of output channels of the CRN encoder to 16 to maintain a lower computational load, while keeping the other configurations identical to the original CRN. In the Spatial Module and Speaker Feature Module, all ConvGLU layers use a kernel size of (2,3). Both the input and output channels for all Spatial Layers are configured to 8. Additionally, the parameter $H$ is set to 8, while $H^{\prime}$ is configured to 32.

We apply the Short-Time Fourier Transform (STFT) using a 20 ms Hann window with a 10 ms shift. At a 16 kHz sampling rate, a 320-point Discrete Fourier Transform (DFT) is used to obtain a 161-dimensional complex spectrum. The model is optimized using the Adam optimizer with an initial learning rate of 0.01, which is halved if the validation loss does not decrease for two consecutive epochs. Training is conducted with a batch size of 16.

% For Short-Time Fourier Transform (STFT), we employ a 20 ms Hann window with a 10 ms shift and use a 320-point Discrete Fourier Transform (DFT) at a 16 kHz sampling rate to obtain a 161-dimensional complex spectrum. The model, optimized with Adam and an initial 0.01 learning rate, reduces the rate by half after two epochs without validation loss improvement. The training batch size is 16.

During online per-frame DOA estimation for the target speaker, silent segments are labeled as non-directional. Consequently, there are 37 possible output classes per frame: 36 directional classes and 1 silence class. Voice Activity Detection (VAD) is applied to the clean target speech to differentiate between silent frames and voiced segments.

We evaluate the performance of DOA estimation using two metrics: 1) Voice Decision Error (VDE), which indicates the percentage of frames misclassified as speech or silence; and 2) Accuracy Rate (AR) \cite{lee2012noise}, which represents the percentage of voiced frames where the estimated DOA falls within a $\pm 10^{\circ}$ range of the ground truth angle.

\subsection{Experiment results}

To the best of our knowledge, prior to our work, only \cite{he20243sicassp} employed voiceprint information to estimate the Direction of Arrival (DOA) of the target speaker, referred to as TS-DOA. Therefore, we scale the TS-DOA method to a parameter level comparable to our proposed system for evaluation purposes.
We provide two versions of RTS-DOA. In the Large version, compared to the standard version, the number of channels in the Speech Enhancement Module is increased to 64, and the parameter $H^{\prime}$ is raised to 96.
Table \ref{tab:noisy} compares the performance of different models at various SIRs between noise and target speech. Table \ref{tab:size} compares the performance of the TS-DOA model and RTS-DOA in terms of parameters and computational cost. Compared to the TS-DOA, both versions of our proposed model show significant improvements in VDE and AR across all SIR levels. This indicates that the proposed methods can more effectively utilize spatial information, thereby enhancing the efficiency of noise processing and more accurately estimating the Direction of Arrival of speech signals, with minimal increase in the consumption of computational resources.

To validate the effectiveness of the proposed DOA estimation framework, beyond the performance improvements attributed to the speech enhancement module, we removed the speech enhancement module and conducted independent tests in ideal conditions (without noise and reverberation). Table \ref{tab:clean} displays the differences in performance and computational resource consumption between different versions of the TS-DOA model and our RTS-DOA model without speech enhancement module. Our DOA estimation framework, while consuming computational resources on par with the compact version of TS-DOA, still maintains a substantial lead in VDE and AR when compared to all models. This is despite the fact that the Huge version of TS-DOA consumes more than ten times the computational resources compared to it. This underscores the superiority of the Spatial Module we use in the task of DOA estimation, demonstrating its effectiveness in utilizing input features and better modeling the spatial information contained within those features.

\begin{table}
  \caption{Comparison of Performance and Computational Resource Consumption Between Different Versions of the TS-DOA Model and RTS-DOA Without Speech Enhance Module}
  \label{tab:clean}
  \centering
  \scalebox{0.8}{
    \begin{tabular}{ c c c c c }
      \toprule
      \multicolumn{1}{c}{\textbf{Metrics}} & 
      \multicolumn{1}{c}{\textbf{VDE} $\downarrow$} &
      \multicolumn{1}{c}{\textbf{AR} $\uparrow$} &
      \multicolumn{1}{c}{\textbf{Params(M)}} & 
      \multicolumn{1}{c}{\textbf{MACs(G/s)}} \\
      \midrule
          TS-DOA(Small) & 0.1167 & 0.6973 & 0.037 & 0.038 \\
          TS-DOA & 0.0769 & 0.8171 & 0.164 & 0.147 \\
          TS-DOA(Huge) & 0.0649 & 0.8555 & 4.609 & 0.730 \\
          \textbf{w/o Enhancement} & \textbf{0.0558} & \textbf{0.9394} & \textbf{0.031} & \textbf{0.065}\\
      \bottomrule
    \end{tabular}
  }
  \vspace{-0.5cm}
  % Remove or place vspace outside the table
\end{table}

Additionally, we conducted ablation studies to validate the effectiveness of other components within the model. Models 1 and 2 in Table \ref{tab:noisy} represent the removal of the Speech Enhancement Module and the Speaker Feature Module from the proposed model, respectively. Comparisons with Model 5 reveal that removing the Speech Enhancement Module results in a noticeable degradation of performance. This indicates the Speech Enhancement Module's critical role in improving the speech signal's clarity and quality before DOA estimation, significantly enhancing the model's ability to accurately estimate the direction of arrival. Similarly, excluding the Speaker Feature Module leads to a decrease in AR(as shown in Model 2 in Table \ref{tab:noisy}), highlighting the importance of the Speaker Feature Module in more effectively utilizing anchor information, thereby aiding in more precise DOA estimation.

Inspired by Hierarchical Representation \cite{He2022HierarchicalSR}, we also attempted to incorporate global features to further utilize speaker information from the anchor(as shown in Model 3 in Table \ref{tab:noisy}). We use a pretrained ECAPA-TDNN \cite{desplanques20_interspeech} as the global feature encoder, following the same structure as described in \cite{ju2022tea}. The output of this network is a 256 dimensional speaker embedding, which we subsequently resize to 1024 dimensions through a trainable linear layer. Afterwards, we multiply this feature with the input of the last Spatial Layer.  However, the inclusion of the global feature resulted in worse experimental outcomes. Upon analysis, this may be due to a misalignment between the feature space encoded by ECAPA-TDNN and the feature space of the hidden states within the DOA estimation module. Besides, we experimented with using the magnitude spectrum instead of complex spectra as the input to the entire system (as shown in Model 4 in Table \ref{tab:noisy}), which resulted in a significant decrease in performance. This observation suggests that the complex spectrum, as opposed to the magnitude spectrum, encompasses more spatial information, rendering it more suitable for DOA estimation tasks.

\section{Conclusions}

% This paper introduces a robust target speaker Direction of Arrival (DOA) estimation system with three modules: Speech Enhancement for improved signal quality, Speaker Feature for optimized anchor use, and Spatial Module for capture spatial information and accurate direction determination. Our approach enhances DOA accuracy in noisy, reverberant environments with interfering and moving speakers, boosting Accuracy Rate (AR) by 0.31 without significant computational demand. Ablation studies validate the distinct contributions of each module to overall performance. 

This paper introduces a robust target speaker direction of arrival estimation system (RTS-DOA). The system utilizes voiceprint information, a jointly optimized speech enhancement module, and advanced spatial modeling to improve DOA estimation accuracy in complex acoustic environments characterized by noise, reverberation, and interfering speakers. Our RTS-DOA system boosts the Accuracy Rate (AR) by 0.31 on the LibriSpeech dataset without significant computational demand, demonstrating its effectiveness in tackling multi-speaker scenarios and establishing new optimal benchmarks.

\bibliographystyle{IEEEtran}
\bibliography{mybib}{}

\begin{thebibliography}{10}
\providecommand{\url}[1]{#1}
\csname url@samestyle\endcsname
\providecommand{\newblock}{\relax}
\providecommand{\bibinfo}[2]{#2}
\providecommand{\BIBentrySTDinterwordspacing}{\spaceskip=0pt\relax}
\providecommand{\BIBentryALTinterwordstretchfactor}{4}
\providecommand{\BIBentryALTinterwordspacing}{\spaceskip=\fontdimen2\font plus
\BIBentryALTinterwordstretchfactor\fontdimen3\font minus \fontdimen4\font\relax}
\providecommand{\BIBforeignlanguage}[2]{{%
\expandafter\ifx\csname l@#1\endcsname\relax
\typeout{** WARNING: IEEEtran.bst: No hyphenation pattern has been}%
\typeout{** loaded for the language `#1'. Using the pattern for}%
\typeout{** the default language instead.}%
\else
\language=\csname l@#1\endcsname
\fi
#2}}
\providecommand{\BIBdecl}{\relax}
\BIBdecl

\bibitem{he20243sicassp}
S.~He, J.~Liu, H.~Li, Y.~Yang, F.~Chen, and X.~Zhang, ``3s-tse: Efficient three-stage target speaker extraction for real-time and low-resource applications,'' in \emph{ICASSP 2024-2024 IEEE International Conference on Acoustics, Speech and Signal Processing (ICASSP)}.\hskip 1em plus 0.5em minus 0.4em\relax IEEE, 2024, pp. 421--425.

\bibitem{schmidt1986multiple}
R.~Schmidt, ``Multiple emitter location and signal parameter estimation,'' \emph{IEEE transactions on antennas and propagation}, vol.~34, no.~3, pp. 276--280, 1986.

\bibitem{roy1989esprit}
R.~Roy and T.~Kailath, ``Esprit-estimation of signal parameters via rotational invariance techniques,'' \emph{IEEE Transactions on acoustics, speech, and signal processing}, vol.~37, no.~7, pp. 984--995, 1989.

\bibitem{huang2001real}
Y.~Huang, J.~Benesty, G.~W. Elko, and R.~M. Mersereati, ``Real-time passive source localization: A practical linear-correction least-squares approach,'' \emph{IEEE transactions on Speech and Audio Processing}, vol.~9, no.~8, pp. 943--956, 2001.

\bibitem{knapp1976generalized}
C.~Knapp and G.~Carter, ``The generalized correlation method for estimation of time delay,'' \emph{IEEE transactions on acoustics, speech, and signal processing}, vol.~24, no.~4, pp. 320--327, 1976.

\bibitem{brandstein1997robust}
M.~S. Brandstein and H.~F. Silverman, ``A robust method for speech signal time-delay estimation in reverberant rooms,'' in \emph{1997 IEEE International Conference on Acoustics, Speech, and Signal Processing}, vol.~1.\hskip 1em plus 0.5em minus 0.4em\relax IEEE, 1997, pp. 375--378.

\bibitem{benesty2004time}
J.~Benesty, J.~Chen, and Y.~Huang, ``Time-delay estimation via linear interpolation and cross correlation,'' \emph{IEEE Transactions on speech and audio processing}, vol.~12, no.~5, pp. 509--519, 2004.

\bibitem{DBLP:journals/tsp/StoicaS90}
\BIBentryALTinterwordspacing
P.~Stoica and K.~C. Sharman, ``Maximum likelihood methods for direction-of-arrival estimation,'' \emph{{IEEE} Trans. Acoust. Speech Signal Process.}, vol.~38, no.~7, pp. 1132--1143, 1990. [Online]. Available: \url{https://doi.org/10.1109/29.57542}
\BIBentrySTDinterwordspacing

\bibitem{tho2014robust}
N.~T.~N. Tho, S.~Zhao, and D.~L. Jones, ``Robust doa estimation of multiple speech sources,'' in \emph{2014 IEEE International Conference on Acoustics, Speech and Signal Processing (ICASSP)}.\hskip 1em plus 0.5em minus 0.4em\relax IEEE, 2014, pp. 2287--2291.

\bibitem{benesty2008microphone}
J.~Benesty, J.~Chen, and Y.~Huang, \emph{Microphone array signal processing}.\hskip 1em plus 0.5em minus 0.4em\relax Springer Science \& Business Media, 2008, vol.~1.

\bibitem{luo2019conv}
Y.~Luo and N.~Mesgarani, ``Conv-tasnet: Surpassing ideal time--frequency magnitude masking for speech separation,'' \emph{IEEE/ACM transactions on audio, speech, and language processing}, vol.~27, no.~8, pp. 1256--1266, 2019.

\bibitem{wang2021multi}
Z.-Q. Wang, P.~Wang, and D.~Wang, ``Multi-microphone complex spectral mapping for utterance-wise and continuous speech separation,'' \emph{IEEE/ACM transactions on audio, speech, and language processing}, vol.~29, pp. 2001--2014, 2021.

\bibitem{luo2020end}
Y.~Luo, Z.~Chen, N.~Mesgarani, and T.~Yoshioka, ``End-to-end microphone permutation and number invariant multi-channel speech separation,'' in \emph{ICASSP 2020-2020 IEEE International Conference on Acoustics, Speech and Signal Processing (ICASSP)}.\hskip 1em plus 0.5em minus 0.4em\relax IEEE, 2020, pp. 6394--6398.

\bibitem{griffiths1982alternative}
L.~Griffiths and C.~Jim, ``An alternative approach to linearly constrained adaptive beamforming,'' \emph{IEEE Transactions on antennas and propagation}, vol.~30, no.~1, pp. 27--34, 1982.

\bibitem{taal2010short}
C.~H. Taal, R.~C. Hendriks, R.~Heusdens, and J.~Jensen, ``A short-time objective intelligibility measure for time-frequency weighted noisy speech,'' in \emph{2010 IEEE international conference on acoustics, speech and signal processing}.\hskip 1em plus 0.5em minus 0.4em\relax IEEE, 2010, pp. 4214--4217.

\bibitem{takeda2016sound}
R.~Takeda and K.~Komatani, ``Sound source localization based on deep neural networks with directional activate function exploiting phase information,'' in \emph{2016 IEEE international conference on acoustics, speech and signal processing (ICASSP)}.\hskip 1em plus 0.5em minus 0.4em\relax IEEE, 2016, pp. 405--409.

\bibitem{xiao2015learning}
X.~Xiao, S.~Zhao, X.~Zhong, D.~L. Jones, E.~S. Chng, and H.~Li, ``A learning-based approach to direction of arrival estimation in noisy and reverberant environments,'' in \emph{2015 IEEE International Conference on Acoustics, Speech and Signal Processing (ICASSP)}.\hskip 1em plus 0.5em minus 0.4em\relax IEEE, 2015, pp. 2814--2818.

\bibitem{chakrabarty2017broadband}
S.~Chakrabarty and E.~A. Habets, ``Broadband doa estimation using convolutional neural networks trained with noise signals,'' in \emph{2017 IEEE Workshop on Applications of Signal Processing to Audio and Acoustics (WASPAA)}.\hskip 1em plus 0.5em minus 0.4em\relax IEEE, 2017, pp. 136--140.

\bibitem{li2018online}
Q.~Li, X.~Zhang, and H.~Li, ``Online direction of arrival estimation based on deep learning,'' in \emph{2018 IEEE International Conference on Acoustics, Speech and Signal Processing (ICASSP)}.\hskip 1em plus 0.5em minus 0.4em\relax IEEE, 2018, pp. 2616--2620.

\bibitem{mack2020signal}
W.~Mack, U.~Bharadwaj, S.~Chakrabarty, and E.~A. Habets, ``Signal-aware broadband doa estimation using attention mechanisms,'' in \emph{ICASSP 2020-2020 IEEE International Conference on Acoustics, Speech and Signal Processing (ICASSP)}.\hskip 1em plus 0.5em minus 0.4em\relax IEEE, 2020, pp. 4930--4934.

\bibitem{mack2022signal}
W.~Mack, J.~Wechsler, and E.~A. Habets, ``Signal-aware direction-of-arrival estimation using attention mechanisms,'' \emph{Computer Speech \& Language}, vol.~75, p. 101363, 2022.

\bibitem{tan2018convolutional}
K.~Tan and D.~Wang, ``A convolutional recurrent neural network for real-time speech enhancement.'' in \emph{Interspeech}, vol. 2018, 2018, pp. 3229--3233.

\bibitem{tan2019learning}
{Tan, Ke and Wang, DeLiang}, ``Learning complex spectral mapping with gated convolutional recurrent networks for monaural speech enhancement,'' \emph{IEEE/ACM Transactions on Audio, Speech, and Language Processing}, vol.~28, pp. 380--390, 2019.

\bibitem{quan2024spatialnet}
C.~Quan and X.~Li, ``Spatialnet: Extensively learning spatial information for multichannel joint speech separation, denoising and dereverberation,'' \emph{IEEE/ACM Transactions on Audio, Speech, and Language Processing}, vol.~32, pp. 1310--1323, 2024.

\bibitem{vaswani2017attention}
A.~Vaswani, N.~Shazeer, N.~Parmar, J.~Uszkoreit, L.~Jones, A.~N. Gomez, {\L}.~Kaiser, and I.~Polosukhin, ``Attention is all you need,'' \emph{Advances in neural information processing systems}, vol.~30, 2017.

\bibitem{he2020speakerfilter}
S.~He, H.~Li, and X.~Zhang, ``Speakerfilter: Deep learning-based target speaker extraction using anchor speech,'' in \emph{ICASSP 2020-2020 IEEE International Conference on Acoustics, Speech and Signal Processing (ICASSP)}.\hskip 1em plus 0.5em minus 0.4em\relax IEEE, 2020, pp. 376--380.

\bibitem{panayotov2015librispeech}
V.~Panayotov, G.~Chen, D.~Povey, and S.~Khudanpur, ``Librispeech: an asr corpus based on public domain audio books,'' in \emph{2015 IEEE international conference on acoustics, speech and signal processing (ICASSP)}.\hskip 1em plus 0.5em minus 0.4em\relax IEEE, 2015, pp. 5206--5210.

\bibitem{dubey2022icassp}
H.~Dubey, V.~Gopal, R.~Cutler, A.~Aazami, S.~Matusevych, S.~Braun, S.~E. Eskimez, M.~Thakker, T.~Yoshioka, H.~Gamper \emph{et~al.}, ``Icassp 2022 deep noise suppression challenge,'' in \emph{ICASSP 2022-2022 IEEE International Conference on Acoustics, Speech and Signal Processing (ICASSP)}.\hskip 1em plus 0.5em minus 0.4em\relax IEEE, 2022, pp. 9271--9275.

\bibitem{allen1979image}
J.~B. Allen and D.~A. Berkley, ``Image method for efficiently simulating small-room acoustics,'' \emph{The Journal of the Acoustical Society of America}, vol.~65, no.~4, pp. 943--950, 1979.

\bibitem{lee2012noise}
B.~S. Lee, \emph{Noise robust pitch tracking by subband autocorrelation classification}.\hskip 1em plus 0.5em minus 0.4em\relax Columbia University, 2012.

\bibitem{He2022HierarchicalSR}
\BIBentryALTinterwordspacing
S.~He, H.~Zhang, W.~Rao, K.~Zhang, Y.~Ju, Y.-R. Yang, and X.~Zhang, ``Hierarchical speaker representation for target speaker extraction,'' 2022. [Online]. Available: \url{https://api.semanticscholar.org/CorpusID:253223893}
\BIBentrySTDinterwordspacing

\bibitem{desplanques20_interspeech}
B.~Desplanques, J.~Thienpondt, and K.~Demuynck, ``{ECAPA-TDNN: Emphasized Channel Attention, Propagation and Aggregation in TDNN Based Speaker Verification},'' in \emph{Proc. Interspeech 2020}, 2020, pp. 3830--3834.

\bibitem{ju2022tea}
Y.~Ju, W.~Rao, X.~Yan, Y.~Fu, S.~Lv, L.~Cheng, Y.~Wang, L.~Xie, and S.~Shang, ``Tea-pse: Tencent-ethereal-audio-lab personalized speech enhancement system for icassp 2022 dns challenge,'' in \emph{ICASSP 2022-2022 IEEE International Conference on Acoustics, Speech and Signal Processing (ICASSP)}.\hskip 1em plus 0.5em minus 0.4em\relax IEEE, 2022, pp. 9291--9295.

\end{thebibliography}

\end{document}